# Investigation of fundamental and higher harmonic AC magnetic susceptibility of FeSe$_{0.5}$Te$_{0.5}$ Superconductor

A. Pal[1,2], P. Rani[1,3], A.K. Hafiz[3], and V. P. S. Awana[1,4,*]

[1]National Physical Laboratory (CSIR) Dr. K. S. Krishnan Road, New Delhi 110012, India
[2]Department of Physics, Indian Institute of Science, Bangalore 560012, India
[3]Centre for Nanoscience and Technology, Jamia Millia Islamia, New Delhi 110025, India
[4]Academy of Scientific and Innovative Research (AcSIR), NPL, New Delhi 110012, India

**Abstract**

We present the complex harmonic magnetic susceptibilities $\chi_n = \chi'_n - i\chi''_n$ (n = 1, 3) of the FeSe$_{0.5}$Te$_{0.5}$ polycrystalline superconducting sample. The *ac* magnetic susceptibility is measured as a function of various external perturbations such as temperature $T$, the ac magnetic field amplitude $H_{ac}$, frequency $\upsilon$, and the magnitude of dc bias field $H_{dc}$. The in-phase ($\chi'_n$) and out of phase ($\chi''_n$) components of the fundamental and third harmonics of ac susceptibility are found to vary as a function of the *ac* drive field. Particularly, the curves shift to lower temperatures with increasing $H_{ac}$. Contrary to ac magnetic field $H_{ac}$, no noticeable change has been observed within the range of the applied dc magnetic field of $H_{dc}$ = 0 Oe to 20 Oe. At a fixed ac field, $H_{ac}$ = 0.5 Oe both parts of the third harmonics show the frequency dependence. The imaginary part of third harmonics, $\chi''_3$ shows two peaks as a function of frequency, one negative in magnitude and another is positive. The negative peak shifts towards the lower temperature and positive peak are shifted towards the higher temperature as we increase the frequency. To better understand the ac magnetic response under the influence of various perturbations, we have analyzed the polar plots (Cole-Cole) of the complex *ac* susceptibility for both the harmonics. Our analysis suggests that the studied sample is in a vortex glass state, characterized by a collective flux creep within Bean's model and the Kim–Anderson model is ruled out.

**Key Words**: Fe Chalcegonide superconductor, *ac* magnetic susceptibility, Higher Harmonics, Cole-Cole plots.

*: *Corresponding Author
Dr. V. P. S. Awana:  E-mail: awana@nplindia.org
Ph. +91-11-45609357, Fax-+91-11-45609310
Homepage awanavps.webs.com

## I. INTRODUCTION

The discovery of superconductivity at moderately high transition temperature ($T_c$ = 26 K) in the LaFeAsO$_{1-x}$F$_x$ ($x$ ~ 0.11) layered compound triggered considerable amount of research activities to look for higher transition temperatures in this class of new materials; being popularly known as Iron-based superconductors [1]. Soon after the discovery, the superconducting transition temperature dramatically exceeded up to 56 K by replacing La with bigger ionic radii rare-earth ion Sm i.e., SmFeAsO$_{1-x}$F$_x$ ($x$ ~ 0.15) [2, 3]. To look for new materials in this class, full substitution of Fe with other 3$d$ metals such as Co was also tried within same crystal structure, which resulted although with nice magnetic properties but not superconducting [4-5]. In the following investigations, another five different families with the same or related crystal structures to that as of superconductors have also been realized, in which iron pnictide FePn (Pn = As, P) or iron-chalcogenide FeCh (Ch= Se, Te) layer was common [6]. The universal existence of Fe-Pn/Ch layer indicates that the active planner iron layer holds the key to high-temperature superconductivity in these materials.

In all the five different families of Fe-based superconducting materials, the binary FeCh-11 type (Ch = S, Se, and Te) system is structurally the simplest (anti-PbO type, space group P4/nmm), one as being composed of only two elements and hence is considered to be the key to hold some of the fascinating physics of unconventional superconductivity in these materials. The FeSe$_{1-\delta}$ in off-stoichiometry form shows superconducting transition at ~ 8 K in ambient environment [7]. The superconducting transition temperature of FeSe increased remarkably from ~ 8 K to 37 K by applying hydrostatic pressure of around ~ 9 GPa [6-9]. Recently, the transition temperature is raised up to 110 K in FeSe monolayer on the STO substrate [10-12]. However, the superconducting transition temperature of bulk FeCh-system is much less compared to their sibling FeAs compounds at ambient conditions, but the simplicity of crystal structure and the high tunability of the transition temperature make them a potential system to understand the mechanism of high-temperature superconductivity in Fe based superconducting materials.

Over the decade, the binary FeSe system has been extensively studied by means of structural, electronic, morphological properties [6, 13], hydrostatic and chemical pressure (doping) [8, 9, 14] and basic mixed state properties [15, 16]. Moreover, due to its high critical current densities, low magnetic field anisotropies and relatively higher value of upper critical field make them attractive materials for the high-power applications at liquid helium temperatures [17–19]. In this regards a quantitative understanding of the nature of their superconducting and mixed state properties such as the pinning properties, the vortex dynamics, and superconducting critical parameters is needed.

The complex magnetic susceptibility is a very sensitive technique to investigate the dynamical magnetic properties of the vortex state and the non-linear processes in flux pinning transport [20-21]. The fundamental and higher harmonics *ac* susceptibility collectively give detailed information regarding different phenomena like the dynamical processes flux flow (FF), thermally assisted flux flow (TAFF) and flux creep [21, 22]; the intragrain and intergrain properties and the non-pinning processes and geometric surface barriers. Bean's Critical state model (CSM) [23, 24] is widely used to investigate the higher harmonics of *ac* susceptibility for type II superconductors. According to this model, the higher harmonics are appeared due to the hysteretic relation between the magnetization and the applied magnetic field caused by the flux pinning. The higher harmonics show different behaviours under various conditions such as variable ac field amplitude, the frequency of *ac* field, *dc* field and temperature. Though, the physical interpretations of the higher harmonics of the *ac* susceptibility as well as the differences in their curve shapes are still needed to be understood very well for various superconductors.

Nevertheless, the critical state model fails to explain some points related to the behaviour of higher harmonics under various perturbations. In this case, it is necessary to consider the influence of linear; thermally assisted flux flow (TAFF), flux flow (FF) and non-linear; flux creep (FC) dynamical dissipative regimes and the properties of different vortex lattice phases [22, 25, 26]. The linearity and non-linearity of the vortex lines diffusion depend upon the interrelation between ac and dc applied a magnetic field. Thus, the higher harmonics of complex susceptibility as a function of various ac and dc magnetic field can provide useful insight about the nature of the dissipative processes, which govern the shape of the Electric field (E)-current (J) characteristics in the superconductors [26, 27].

In this paper, we report a comprehensive study of the complex magnetic susceptibility of the FeSe$_{0.5}$Te$_{0.5}$ polycrystalline superconducting sample. The fundamental and third harmonics of the *ac* complex susceptibility has been studied as a function of the temperature $T$ ($2 \leq T \leq 15$ K), the ac magnetic field amplitude $H_{ac}$ (0.2 Oe $\leq H_{ac} \leq$ 5 Oe), frequency υ (333 Hz ≤ υ ≤ 9999 Hz), and the magnitude of a superimposed *dc* field $H_{dc}$ (0 Oe ≤ $H_{dc}$ ≤ 20 Oe) to understand the vortex dynamics and granularity effects of the superconducting and mixed state. The peak position of the imaginary part of complex *ac* susceptibility shifted towards to lower temperatures with increasing the applied *ac* magnetic field. The superconducting transition temperature determined by *dc* magnetic field does not have much influence of the *ac* - *dc* magnetic field and frequency. Our detailed *ac* susceptibility results in terms of the variation of various harmonics with applied bias ac field and variation in frequency indicate that studied Fe chalcogenide superconductor follows the vortex glass state within Bean's model and the Kim–Anderson model is ruled out.

**II. EXPERIMENTAL METHODS**

The polycrystalline $FeSe_{0.5}Te_{0.5}$ sample is synthesized by standard solid-state reaction route using high purity 3N Fe, Se and Te in their elemental form. The precursors were weighed in their stoichiometric ratio and then ground thoroughly using mortal pestle in a controlled atmosphere. The mixed material pressed into a pellet and encapsulated in an evacuated ($10^{-3}$ Torr) quartz tube and then heated at 750°C for 12 hours. After that, the furnace is allowed to cool down to room temperature slowly. The X-ray diffraction (XRD) are taken on Rigaku diffractometer with Cu-$K_\alpha$ (λ = 1.54 Å) to check the phase purity of the sample. The crystal structure is refined by the Rietveld method using open source FullProf program. No other extra peaks of any parasitic phases are observed in the fitted data, so the sample is considered as homogeneous and pure in phase. The details of the sample preparation and the magnetic and transport properties are reported somewhere [14].

The ac-susceptibility measurements were carried out using a Quantum Design AC Measurement System (ACMS) option for the Physical Property Measurement System (PPMS). The sample in the form of a rectangular slab was mounted on the experimental setup and measured its magnetic response under an applied alternating *ac* and *dc* magnetic fields. The measurement has been carried out at the various frequencies (υ = 333 to 9999 Hz), and the amplitude (0.0 to 5 Oe) of the ac applied a magnetic field.

## III. RESULTS AND DISCUSSION

### A. Phase purity (X-Ray diffraction)

Fitted and observed room temperature powder X-ray diffraction (PXRD) patterns of studied FeSe$_{0.5}$Te$_{0.5}$ sample are shown in Fig. 1. It is evident from this figure 1 that the studied sample is free of any impurities within the XRD limit. The sample is crystallized in P4/nmm space group in tetragonal structure, the Rietveld refined lattice parameters are a = b = 3.800(1) Å and c = 6.019(3) Å. The lattice parameters are in good agreement with earlier reports on similar samples [7-9, 14-16]. The inset of figure 1 shows the *ac* susceptibility of the studied sample at amplitude of 0.5 Oe and frequency 333 Hz. The real part clearly shows the diamagnetic transition at around 9.5 K coupled with the peak in imaginary part starting from the same temperature. It is clear from figure 1 that the studied sample is a bulk superconductor below 9.5 K.

### B. AC magnetic response at different $H_{ac}$ amplitudes

Figure 2(a) and (b) depict the real ($\chi'_1$), and imaginary ($\chi''_1$) component of the fundamental and third harmonics of complex magnetic susceptibility measured at 333 Hz as a function of temperature under different *ac* applied magnetic field $H_{ac}$ = 0.2 Oe - 5 Oe for studied FeSe$_{0.5}$Te$_{0.5}$ sample. The *dc* bias field is zero, $H_{dc}$ = 0 Oe throughout the measurement. The in-phase (real) component of complex magnetic susceptibility, shows the superconducting transition temperature $T_c$ ~ 9.5 K, which also coincides with the first positive signal of the imaginary component $\chi''_1$ (shown in the inset of figure 1(a)). Figure 2(a) clearly indicates that although the value of $T_c$ is independent of the amplitude of the *ac* magnetic field, the diamagnetic signal at 2 K increases. On the other hand, the peak temperature, $T_p$ (shown in inset of figure 2(a)) of $\chi''_1$ is sharply decreased with increasing amplitude of *ac* applied field, $H_{ac}$. Both curves (see figure 1(a)) are strongly influenced by the increase of the *ac* field amplitude. The absolute value of $\chi'_1$ is directly proportional to applied *ac* field amplitude. Correspondingly, the height of the peak in the imaginary part (inset of Figure 2(a)) increases with increasing $H_{ac}$. The $T_p$ is linearly decreased from 8.5 K to 2.4 K as $H_{ac}$ increases from 0.2 to 5 Oe.

Generally, high $T_c$ superconductors show two different peaks in the imaginary part of complex susceptibility, which are accounted for the inter- and intra-granular transitions. In the currently studied sample, we do not observe any separate peak at lowest applied *ac* field also [28]. Thus, we can neglect the effect of granularity in the sample, or one can say that due to the strong inter-granular coupling, the transition from inter- and intra-grain components merge into each other. Consequently, a single diamagnetic transition is found. All the curves expectedly shift towards lower temperatures with increasing magnetic field amplitude ($H_{ac}$), owing to the easier penetration of the flux into the sample. The shift in the peak with increasing field is accompanied by a considerable increase in the height of the peak, especially for the highest amplitude. The third harmonics show (see figure 2(b)) opposite behaviour to the as reported for $Fe_{1.02}Se$ but somewhat similar to $LaFeAsO_{0.92}F_{0.08}$ compound [15, 29]. The real part of third harmonic shows a single negative peak, whereas the imaginary part shows two different peaks. All the peaks in real and imaginary parts become wider and shift towards lower temperature by increasing $H_{ac}$. In the real part, the peak grows in a negative direction with a noticeable height. However, in the case of the imaginary part, the peak near to $T_c$ shifts toward a positive direction on increasing field amplitude.

The Cole-Cole plots, $\chi''_1(\chi'_1)$ for fundamental and third harmonics obtained from figure 2 at different *ac* applied field amplitudes for studied $FeSe_{0.5}Te_{0.5}$ sample are shown in figure 3. As evident from figure 3, the $\chi''_1(\chi'_1)$ plots are strongly influenced by the applied *ac* field amplitude. The peak height and the broadness of the first harmonic increases with increasing ac applied field. In Cole-Cole plots also we have observed only one peak, which again shows that the intra-grain and inter-grain contributions merge together. The third harmonics of $FeSe_{0.5}Te_{0.5}$ (shown in Figure. 3(b)) shows entirely different shape in comparison to the one as found in fundamental *ac* susceptibility of polycrystalline superconductors. It forms a closed contour with an ellipsoidal (lens) shape. The closed contour is mostly situated in the III quadrant, with a tiny portion placed in the neighbouring II and IV quadrants. No part of the curve is seen in I quadrant. Applied field amplitude triggered a visible change in the behaviour of the curve; area of the curve considerably increases with increasing field. In particular, at highest amplitude, the curve occupies the large area as compared to the low field amplitude. The general behaviour of Cole-Cole plots is similar to that as reported for another superconductor NdFeAsO/F [29].

**C. AC magnetic response at different frequencies**

The temperature dependence of the fundamental and third harmonic of *ac* complex magnetic susceptibility at representative frequencies for studied FeSe$_{0.5}$Te$_{0.5}$ sample is shown in figure 4(a). The data for both the harmonics were collected at a fixed amplitude of *ac* field, $H_{ac}$ = 0.5 Oe and zero bias dc field. The in-phase component ($\chi'_1$) of first harmonic shows only one peak as a function of temperature which indicates a strong inter-granular coupling. This behaviour is different from the general character high-$T_c$ superconductors where inter- and intra-grain peaks can be observed separately. Figure 4(a) shows that the diamagnetic transition of the in-phase component of susceptibility (main panel of figure) and the temperature of the peak of out of phase component of first harmonics, $T_p$, shifts towards the higher temperature with increasing the frequency from 333 Hz to 9999 Hz. It was shown numerically that in both the case, vortex glass as well as Kim–Anderson, the diamagnetic transition of $\chi'_1$ and peak position of $\chi''_1(T_p)$ shift towards the higher temperature with increasing frequency [31]. It is the height of the $\chi''_1(T_p)$ peak which distinguishes both the scenarios. The $\chi''_1(T_p)$ peak increases in a vortex glass state whereas decreases in Kim-Anderson case, in which the vortices are still in a liquid phase (although highly viscous). It is clear from the inset of figure 4(a) that the height of the peak $\chi''_1(T_p)$ grows as we increase the frequency which indicates that the studied system is in a Vortex glass phase. A similar splitting effect is also observed in polycrystalline Fe$_{1.02}$Se and LaFeAsO$_{0.92}$F$_{0.08}$ samples [14, 28].

We also performed a similar analysis on the 3$^{rd}$ harmonics of the *ac* magnetic susceptibility for studied FeSe$_{0.5}$Te$_{0.5}$ sample. Figure 4(b) shows the real and imaginary parts of the third harmonics of the studied sample at various frequencies. The frequency dependence of the third harmonics of FeSe$_{0.5}$Te$_{0.5}$ is similar to the previously reported LaFeAsO$_{0.92}$F$_{0.08}$ superconductor [28]. It is obvious from figure 4(b) that 3$^{rd}$ harmonic components are even more sensitive to the frequency variation. At low frequency, the real part shows a negative peak (closer to $T_c$) followed by a small but wide positive hump. Further, it shows a reduction of the negative peak as the frequency increases, followed by a rise in the height and the width of the following maximum. The rapid change from negative to positive values indicate that the system is growing from a stable pinning state (critical state) at the lowest frequencies to the domination of the dissipative regimes at high frequencies [22, 25]. Both the negative and positive peaks of the real component of 3$^{rd}$ harmonics do shift towards higher temperatures with an increase in frequency.

The out of phase component ($\chi''_3$) peak shows a reduction in peak height and shifts toward the higher temperature as we increase the frequency. The imaginary component is also strongly influenced by varying frequency. Initially, a small positive peak (maximum) near to transition temperature is found which is further followed by a subsequent deeper negative peak (minimum) at a lower temperature. The positive peak diminishes with increasing frequency and completely disappears at the highest frequency (9999 Hz), together with a constant increase in depth of the following minimum. The obtained curve is different from the Bean critical state geometry [31].

Figure 5(a) illustrates the cole-cole plots of the first harmonics for studied FeSe$_{0.5}$Te$_{0.5}$ sample which also support the above analysis. Due to strong inter-granular coupling, the inter-grain and the intra-grain contributions merge into each other. The peak height of the dome-shaped curves is directly proportional to the frequency, which is contrary to the Kim–Anderson model [31]. The growth in the height of the maximum of polar (Cole-Cole) plots (see Fig. 5(a)) and the shift towards the higher temperature in the peak of imaginary signal (See Fig. 4(b)) as a function of frequency is consistent with the simulated vortex glass collective creep model [31].

The Cole-Cole plots of third-harmonic as a function of frequency are shown in figure 5(b). As evident from the figure, the polar plots for the lowest frequency of 333 Hz are almost entirely in the left half. The cole-cole plots shift towards the right half quadrant with increasing frequency. The area of plots is somewhat unchanged with frequency, this contrary to the impact of *ac* driven field (fig. 3(b)). The cole-cole plots at higher frequencies appear to behave somewhat like the Bean curves, which are closed loops, all staying in 3$^{rd}$ and 4$^{th}$ quadrants. Although the detailed *ac* susceptibility results in terms of the variation of various harmonics with applied bias *ac* field and variation in frequency indicate the Fe chalcogenide superconductor to follow the vortex glass state within Bean's model and the Kim–Anderson model is ruled out, yet some further studies are needed for a complete understanding of the material [30]. Some of the preliminary findings of this study were reported by us in a symposium, but without thorough analysis [32]

To understand the magnetic response of the studied sample more deeply the analysis of frequency dependence of $\chi''$ peak was performed using Vogel-Fulcher law. According to Vogel-Fulcher law the frequency dependent spin freezing temperature, $T_f$ (temperature of the $\chi''$ peak) can be described by [33]

$$\omega = \omega_0 \exp\left(\frac{-E_a}{K_B (T_f - T_0)}\right) \quad (1)$$

Where $E_a$ is the activation energy for the relaxation process, $\omega_0$ is the characteristic frequency of the clusters, and $T_0$ is the Vogel-Fulcher temperature, which is the measure of inter-cluster interaction strength. The phenomenological Vogel-Fulcher (VF) law takes accounts of the interaction of magnetic clusters. In no interaction scenarios, i.e., $T_0 = 0$, the Vogel-Fulcher equation is transformed to the Arrhenius equation [35, 36] which describes the relaxation processes of non-interacting magnetic clusters as,

$$\omega = \omega_0 \exp\left(\frac{-E_a}{K_B T_f}\right) \quad (2)$$

The Vogel-Fulcher law fit to the experimental data of the $FeSe_{0.5}Te_{0.5}$ sample is shown in figure 6. It is evident from the figure (6) that the freezing temperature $T_f$ and $1/\ln(\omega_0/\omega)$ follows the expected linear behaviour. From the best linear fit, we obtained $\omega_0 = 10^{12}$ Hz, $E_a = 52.65 \pm 3$K, and $T_0 = 8.69$ K. The value of the $T_0 = 8.69$ K is in agreement with the value of freezing temperature $T_f = 9.13$ K, obtained from the $ac$ susceptibility measurements. Hence, the fit of the experimental data of Vogel-Fulcher law indicates clearly the presence of a spin-glass state in the studied sample.

**D. AC magnetic susceptibility response at different $dc$ biased ($H_{dc}$) fields**

To gain more insight of the vortex dynamics of the studied $FeSe_{0.5}Te_{0.5}$ sample, we have also measured the first and third harmonics at the different $dc$ biased magnetic field, $H_{dc} = 5$ Oe, 10 Oe, and 20 Oe (shown in figure 7(a)) and 7(b)). The $\chi'_1(T)$ shows a usual transition which is also shown in $dc$ susceptibility. We want to emphasize that unlike to the first harmonic of $ac$ magnetic susceptibility under varying $ac$ magnetic field amplitudes, the $dc$ applied field does not show any noticeable effect on both the real and imaginary part (inset of figure 7(a)). All the curves show the same diamagnetic transition and $\chi'_1(T)$ grows towards low temperature in the negative direction. The same behaviour is followed in the imaginary component $\chi''_1(T)$ as the peak position under any $dc$ magnetic field remains unaffected.

Figure 7(b) shows the real component of the 3$^{rd}$ harmonic curve $\chi'_3(T)$, having large initial minima, which remains same on increasing field amplitude $H_{dc}$. One can say that it is unaffected by the applied *dc* magnetic field. Each curve is dominated by the following small maximum, which remains same on increasing field amplitude. The inset of figure 7(b) shows that for the increasing $H_{dc}$, the small maxima (closer to $T_c$) of $\chi''_3(T)$ visually remain unchanged. Each curve is dominated by the following deeper minimum, which also remains the same with increasing $H_{dc}$.

The polar plot (Cole-Cole) of fundamental and third harmonic obtained from the curves of figure 7(a) and 7(b) are shown in figure 8(a) and 8(b), respectively. The closed contour is mostly situated in the III quadrant, with a tiny portion placed in the neighbouring II and IV quadrants. There is a no visible change in the behaviour triggered by increase in $H_{dc}$.

## IV. CONCLUSION

In conclusion, the fundamental and third harmonics of the complex magnetic susceptibility has been studied for FeSe$_{0.5}$Te$_{0.5}$ superconducting polycrystalline samples as functions of the temperature *T*, at various amplitudes, $H_{ac}$, and frequency, υ of *ac* magnetic field, and the magnitude of dc bias field $H_{dc}$. The studied sample shows strong inter-granular coupling. Our analysis suggests that sample is in a vortex glass state, characterized by a collective flux creep. To better understand the ac magnetic response under the influence of various perturbations we have analyzed the polar plot (Cole-Cole) of the complex *ac* susceptibility for both the harmonics.

## ACKNOWLEDGMENTS

We would like to thank Director NPL for his keen interest in this work. Poonam Rani would like to acknowledge CSIR for providing SRF as a financial assistance.

**Figure Captions**

Figure 1 (Colour on line) Observed and fitted powder X-ray diffraction (PXRD) patterns of studied FeSe$_{0.5}$Te$_{0.5}$ sample, inset shows the *ac* susceptibility of the same at amplitude of 0.5 Oe at 333 Hz.

Figure 2 (Colour online) Real ($\chi'_1$) and imaginary ($\chi''_1$) (inset) component of (a) fundamental and (b) third harmonics of *ac* magnetic susceptibility for FeSe$_{0.5}$Te$_{0.5}$ measured at various applied ac field $H_{ac}$ = 0.2 Oe, 0.5 Oe, 0.7 Oe, 1 Oe, 2 Oe and 5Oe as a function of temperature. For this measurement the frequency, $\upsilon$ = 333 Hz was fixed and no DC field was present $H_{ac}$= 0.0 Oe. The solid lines through the points are to guide the eye only.

Figure 3 (Colour online) The Cole-Cole plots, $\chi''(\chi')$ of FeSe$_{0.5}$Te$_{0.5}$ for (a) fundamental and (b) third-harmonic, measured at different ac drive field H$_{ac}$ = 0.2 Oe, 0.5 Oe, 0.7 Oe, 1Oe, 2 Oe and 5 Oe at fixed frequency $\upsilon$ = 333 Hz and $H_{dc}$ = 0.0 Oe. The solid lines are a guide for the eyes.

Figure 4 (Colour online) Real ($\chi'_1$) and imaginary ($\chi''_1$) (inset) component of (a) fundamental and (b) third harmonics of ac magnetic susceptibility vs temperature at representative frequencies, $\upsilon$ = 333 Hz, 999 Hz, 3333 Hz, 6666 Hz and 9999 Hz. The data have been collected at $H_{dc}$ = 0.0 Oe and $H_{ac}$ = 0.5 Oe constant field values. The solid lines are a guide for the eyes.

Figure 5 (Colour online) (a) Fundamental and (b) third harmonic cole – cole plots $\chi''(\chi')$ for studied FeSe$_{0.5}$Te$_{0.5}$ at representative frequencies, $\upsilon$ = 333 Hz, 999 Hz, 3333 Hz, 6666 Hz and 9999 Hz. The measurement has been done at a fixed amplitude of $H_{ac}$ = 0.5 Oe of the ac magnetic field and zero dc biased field. The solid lines are a guide for the eyes.

Figure 6 (Colour online) Variation of the freezing temperature, $T_f$ with the frequency of the *ac* field in a Vogel-Fulcher plot. The solid line is the best fit of Eq. (1).

Figure 7 (Colour online) Temperature dependence of real ($\chi'$) and imaginary ($\chi''$) (inset) part of (b) fundamental and (b) third harmonic of *ac* magnetic susceptibility at representative *dc* magnetic field $H_{dc}$ = 5 Oe, 10 Oe and 20 Oe. The measurement have been done at fixed frequency $\upsilon$ = 333 Hz and fixed ac drive field $H_{ac}$ = 0.5 Oe. The solid lines through the data are drawn to guide the eyes.

Figure 8 (Colour online) The Cole–Cole plots for the (a) first and (b) third harmonics, $\chi'_3$ ($\chi''_3$) of ac magnetic susceptibility of FeSe$_{0.5}$Te$_{0.5}$, measured at different dc magnetic field $H_{dc}$ =5 Oe, 10 Oe and 20 Oe as a function of temperature with constant ac drive field $H_{ac}$ = 0.5 Oe and frequency $\upsilon$ = 333 Hz. The solid lines through the data are drawn to guide the eyes.

Figure 1

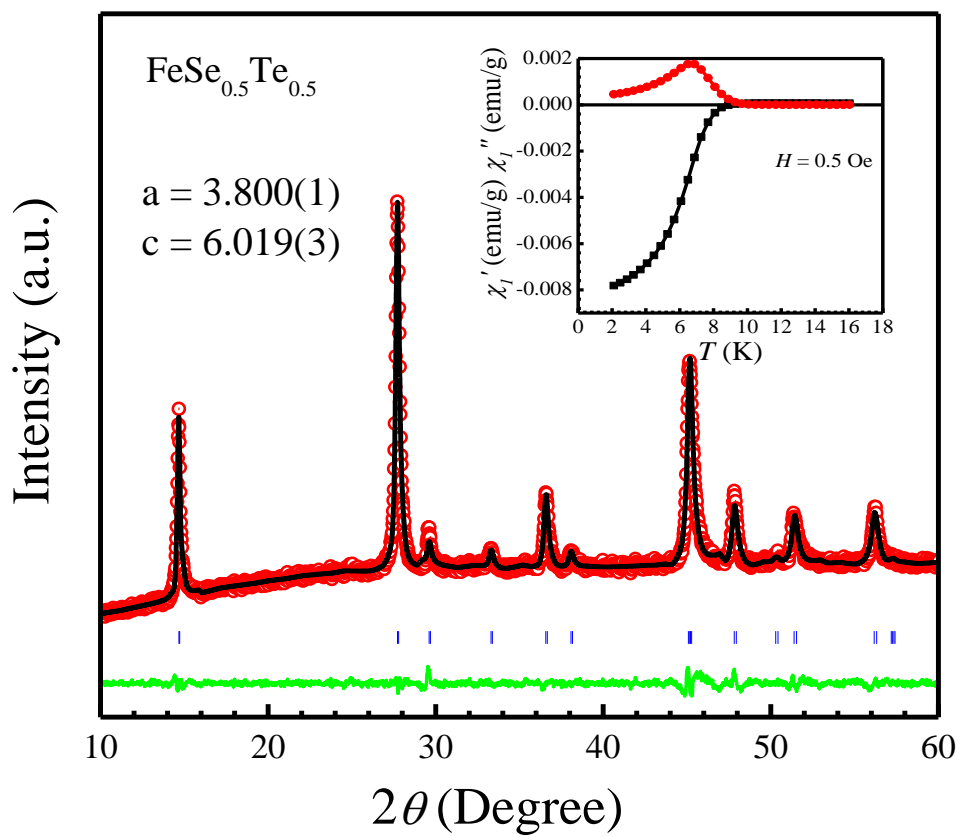

Figure 2

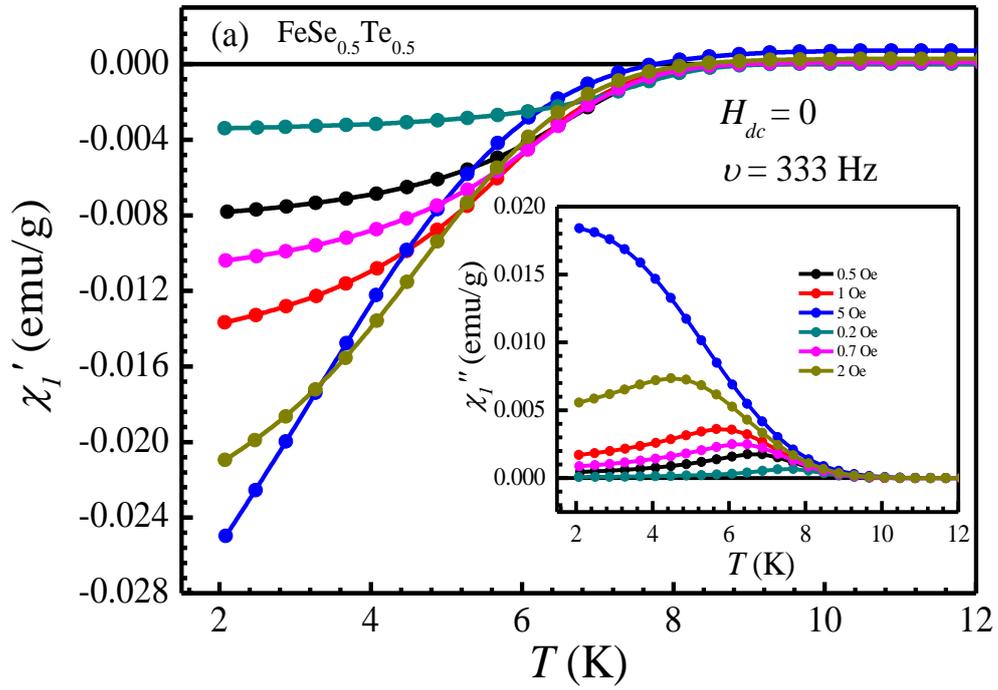

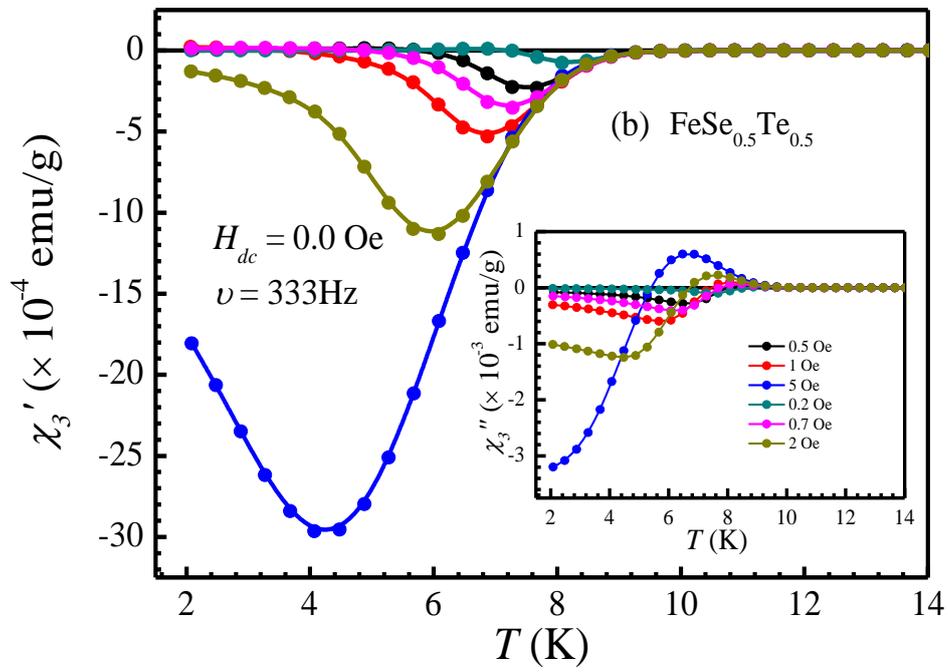

Figure 3

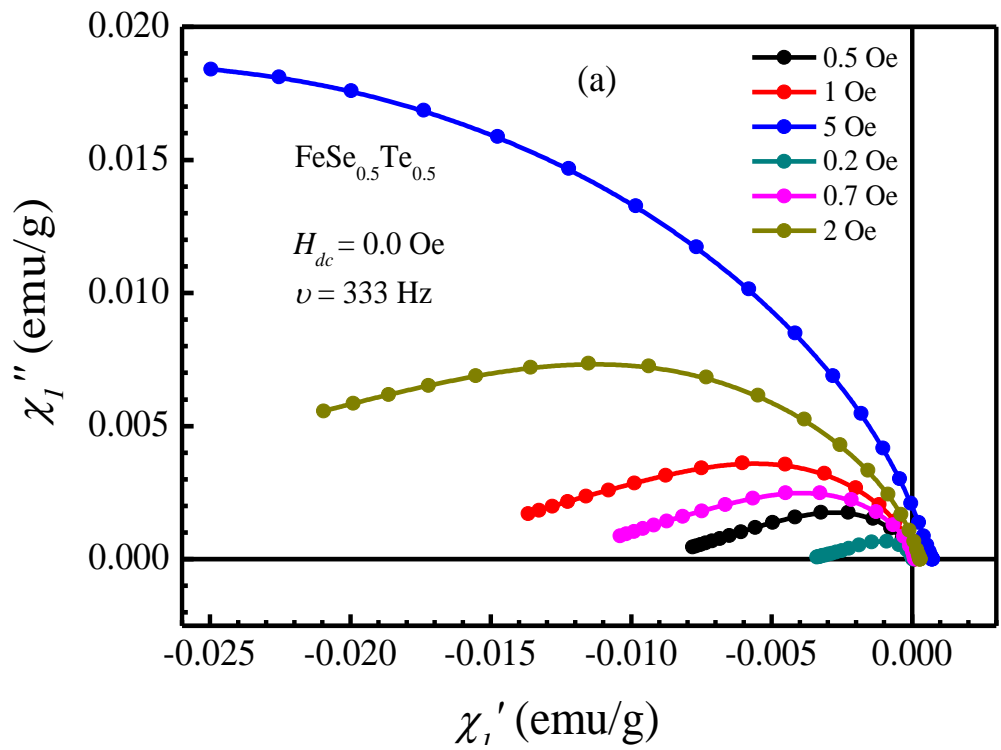

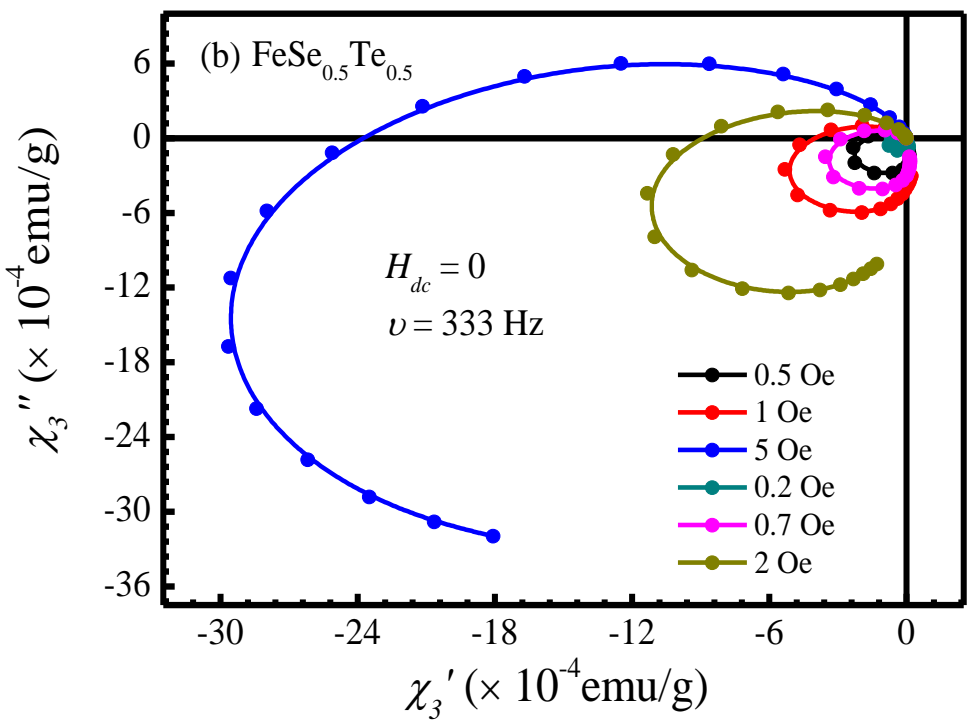

Figure 4

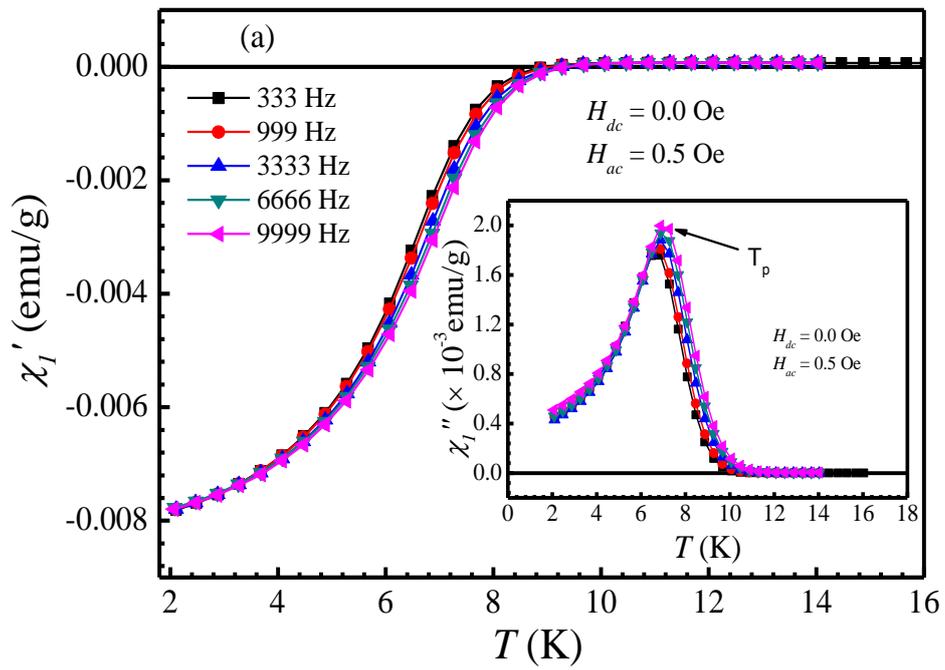

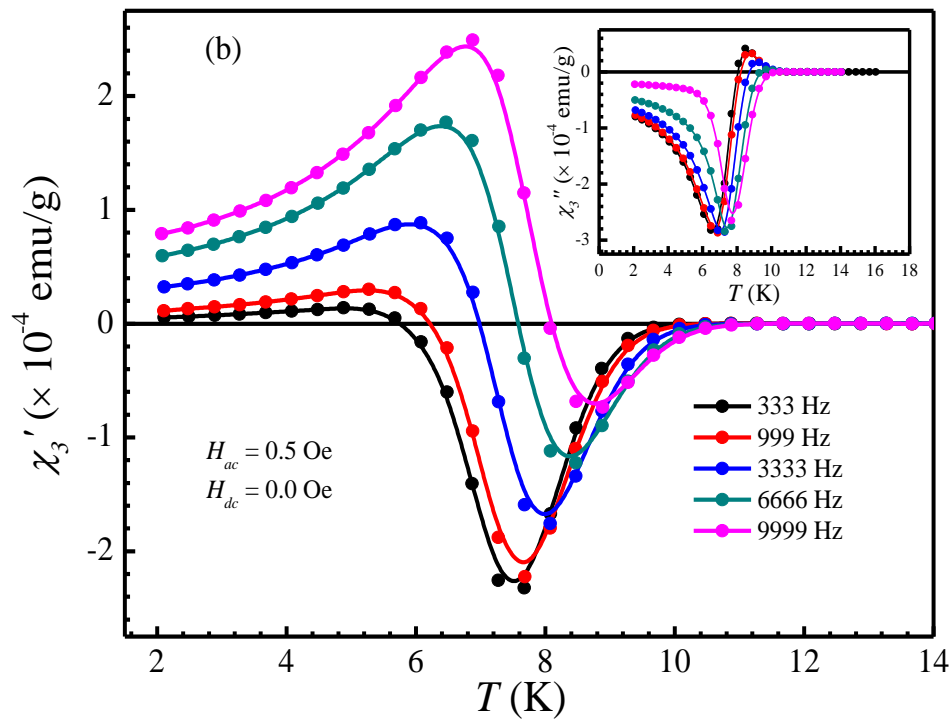

Figure 5

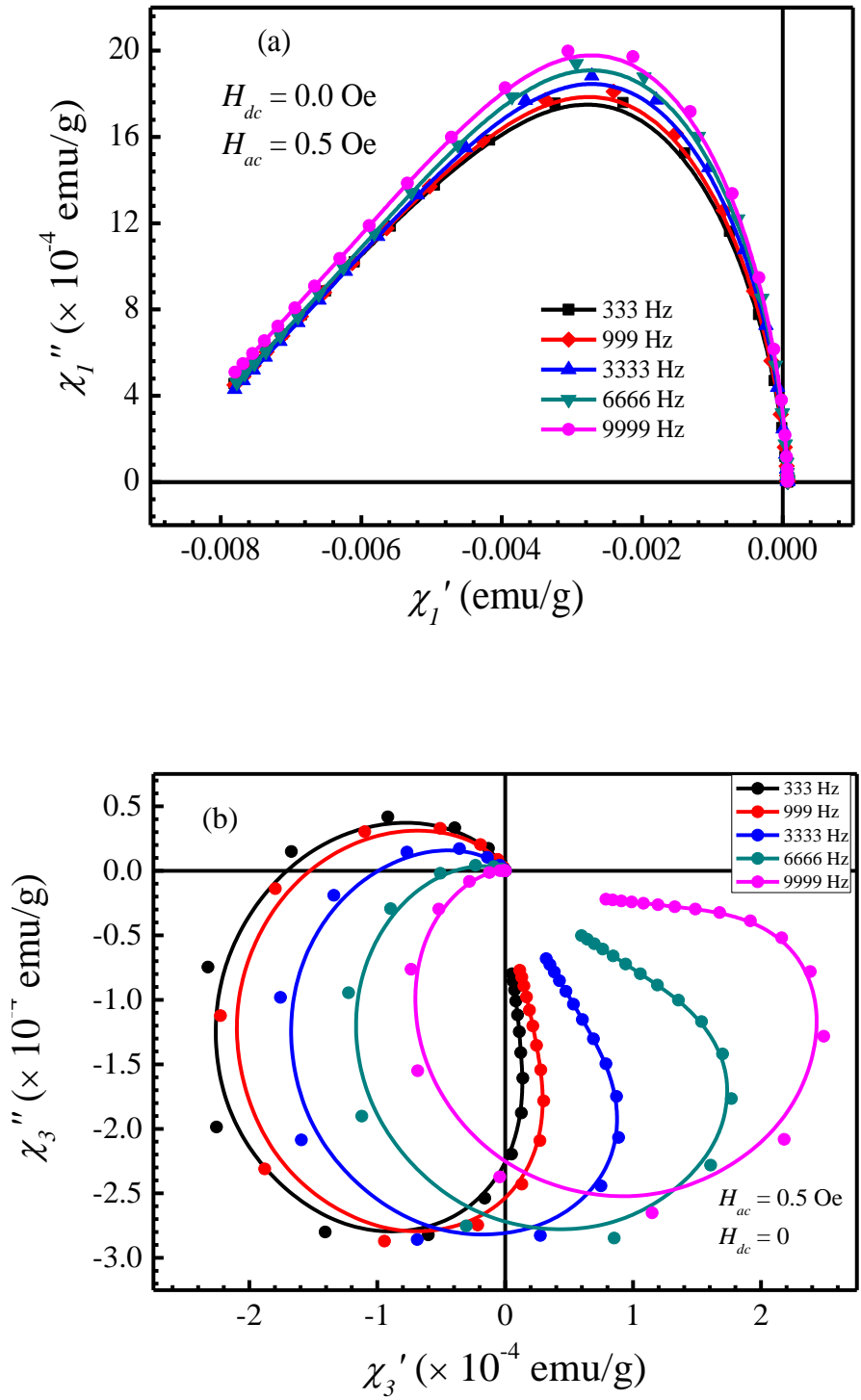

Figure 6

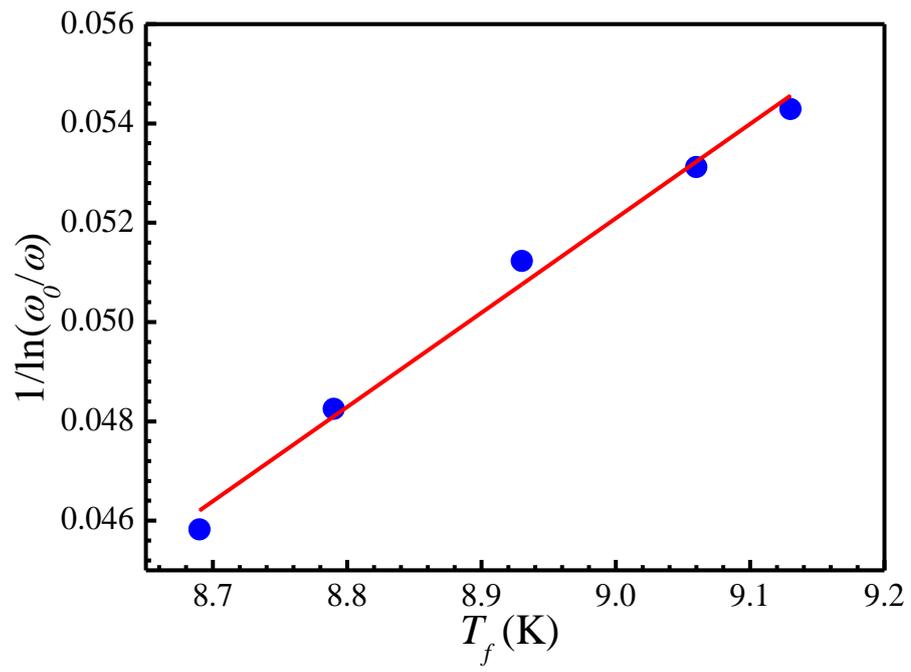

Figure 7

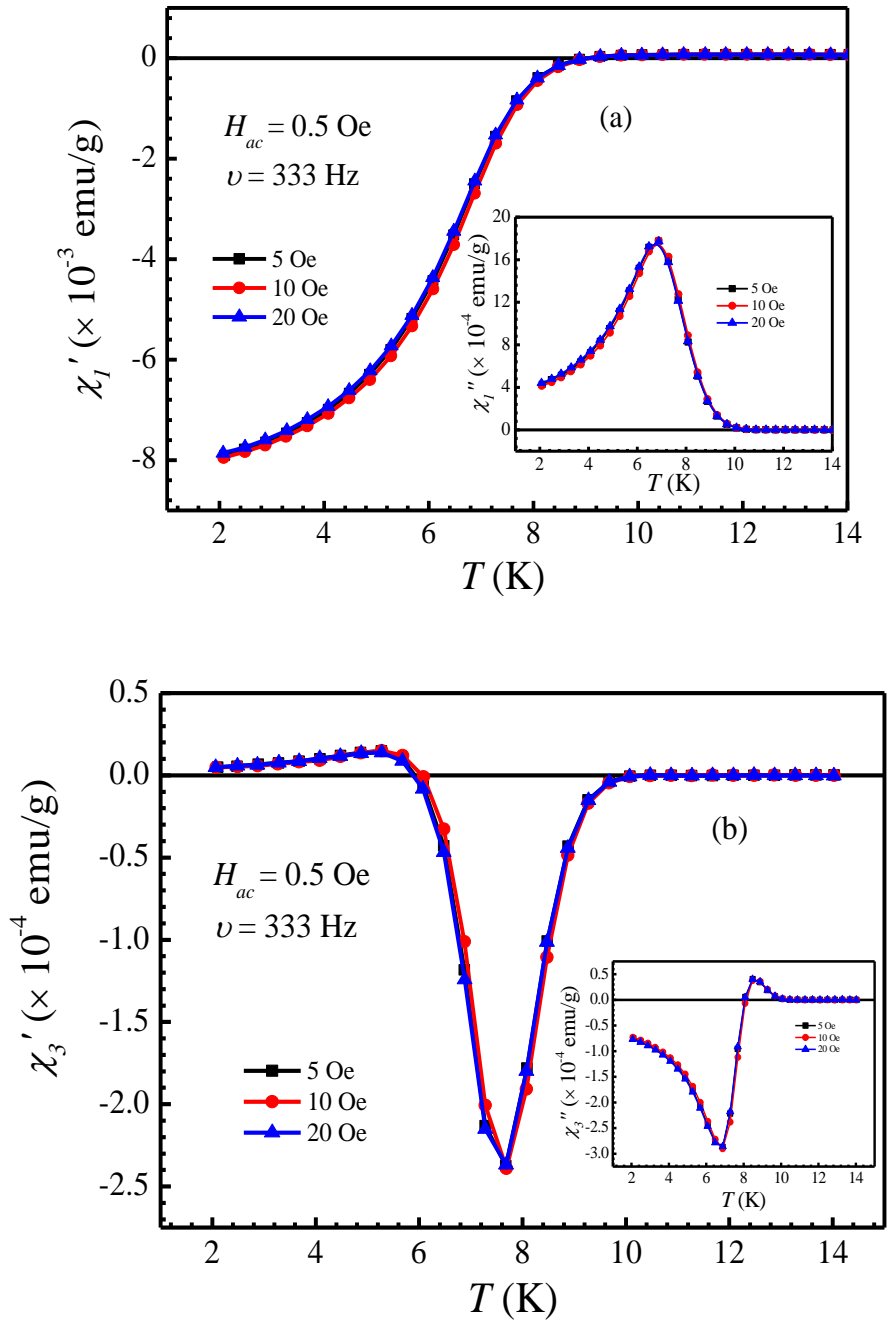

Figure 8

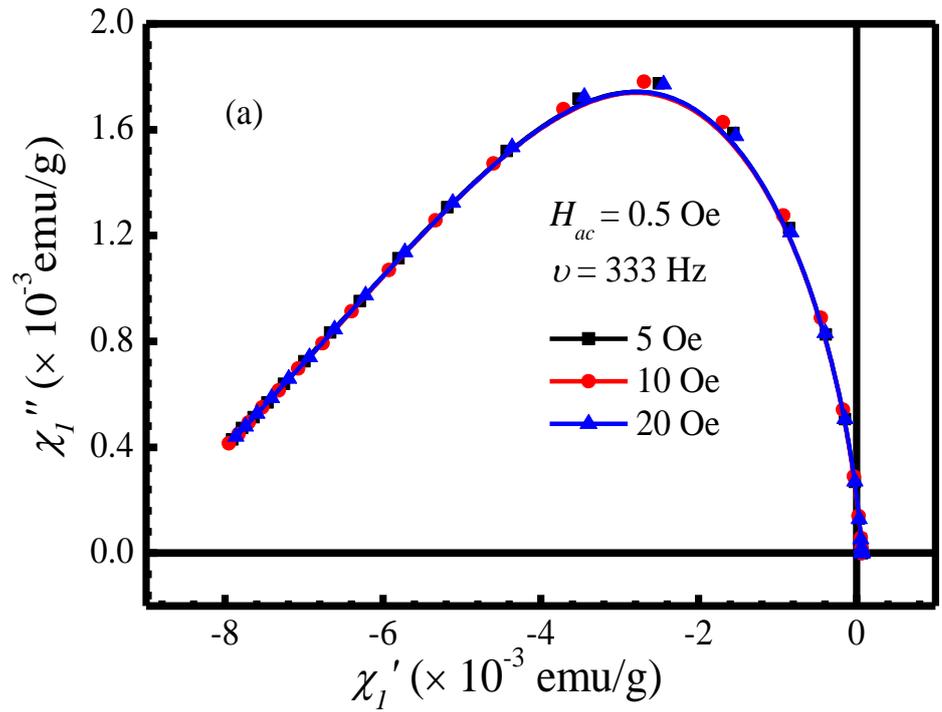

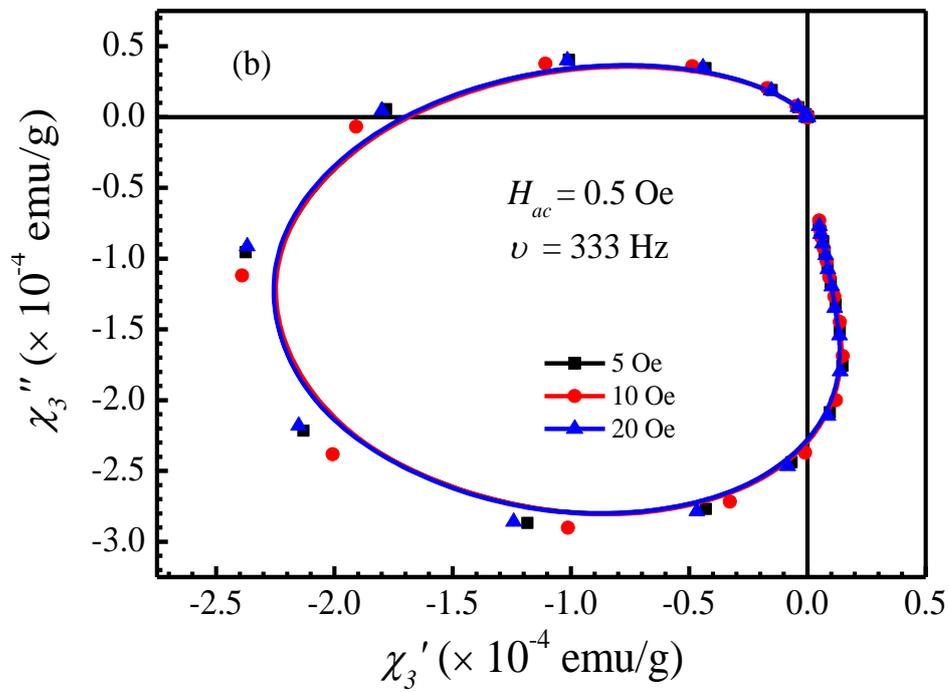